\documentclass[sigplan,screen,nonacm]{acmart}

\usepackage{booktabs} % For formal tables

\usepackage{agda}
\usepackage{catchfilebetweentags}
\usepackage{balance}

\newcommand{\AK}{\AgdaKeyword}

\newcommand{\AStr}{\AgdaString}
\newcommand{\AN}{\AgdaNumber}
\newcommand{\AD}{\AgdaDatatype}
\newcommand{\AF}{\AgdaFunction}
\newcommand{\AR}{\AgdaRecord}
\newcommand{\ARF}{\AgdaField}
\newcommand{\AB}{\AgdaBound}
\newcommand{\AIC}{\AgdaInductiveConstructor}

\usepackage[utf8]{inputenc}
\usepackage{newunicodechar}
\newunicodechar{′}{\ensuremath{\prime}}
\newunicodechar{−}{\ensuremath{-}}
\newunicodechar{─}{\ensuremath{-}}
\newunicodechar{◆}{\ensuremath{\Diamondblack}}
\newunicodechar{⧫}{\ensuremath{\blacklozenge}}
\newunicodechar{∷}{\ensuremath{::}}
\newunicodechar{∙}{\ensuremath{\bullet}}
\newunicodechar{□}{\ensuremath{\Box}}
\newunicodechar{∎}{\ensuremath{\blacksquare}}
\newunicodechar{⋆}{\ensuremath{\star}}

\newunicodechar{₀}{\ensuremath{_0}}
\newunicodechar{₁}{\ensuremath{_1}}
\newunicodechar{₂}{\ensuremath{_2}}
\newunicodechar{₃}{\ensuremath{_3}}

\newunicodechar{ₖ}{\ensuremath{_k}}
\newunicodechar{ₘ}{\ensuremath{_m}}
\newunicodechar{ₙ}{\ensuremath{_n}}
\newunicodechar{ᵣ}{\ensuremath{_r}}
\newunicodechar{ₛ}{\ensuremath{_s}}

\newunicodechar{₊}{\ensuremath{_+}}

\newunicodechar{⁺}{\ensuremath{\textsuperscript{+}}}
\newunicodechar{⁻}{\ensuremath{\textsuperscript{-}}}

\newunicodechar{²}{\ensuremath{^2}}

\newunicodechar{ⁱ}{\ensuremath{^i}}
\newunicodechar{ˡ}{\ensuremath{^l}}
\newunicodechar{ʳ}{\ensuremath{^r}}

\newunicodechar{ᴬ}{\ensuremath{^A}}
\newunicodechar{ᴮ}{\ensuremath{^B}}
\newunicodechar{ᴵ}{\ensuremath{^I}}
\newunicodechar{ᴿ}{\ensuremath{^R}}
\newunicodechar{ᵀ}{\ensuremath{^T}}
\newunicodechar{ᵁ}{\ensuremath{^U}}
\newunicodechar{ⱽ}{\ensuremath{^V}}

\newunicodechar{⋯}{\ensuremath{\cdots}}

\newunicodechar{≡}{\ensuremath{\equiv}}
\newunicodechar{≢}{\ensuremath{\not\equiv}}
\newunicodechar{≟}{\mbox{\tiny\ensuremath{\stackrel{?}{=}}}}
\newunicodechar{≈}{\ensuremath{\approx}}

\newunicodechar{≤}{\ensuremath{\le}}
\newunicodechar{≥}{\ensuremath{\ge}}

\newunicodechar{→}{\ensuremath{\rightarrow}}
\newunicodechar{←}{\ensuremath{\leftarrow}}
\newunicodechar{⇒}{\ensuremath{\Rightarrow}}
\newunicodechar{⇉}{\ensuremath{\rightrightarrows}}

\newunicodechar{∂}{\ensuremath{\partial}}
\newunicodechar{∋}{\ensuremath{\ni}}
\newunicodechar{∞}{\ensuremath{\infty}}
\newunicodechar{∀}{\ensuremath{\forall}}
\newunicodechar{∃}{\ensuremath{\exists}}
\newunicodechar{⊢}{\ensuremath{\vdash}}
\newunicodechar{⟨}{\ensuremath{\langle}}
\newunicodechar{⟩}{\ensuremath{\rangle}}
\newunicodechar{⊤}{\ensuremath{\top}}
\newunicodechar{∘}{\ensuremath{\circ}}
\newunicodechar{⊎}{\ensuremath{\uplus}}
\newunicodechar{×}{\ensuremath{\times}}
\newunicodechar{ℕ}{\ensuremath{\mathbb{N}}}
\newunicodechar{⟦}{\ensuremath{\llbracket}}
\newunicodechar{⟧}{\ensuremath{\rrbracket}}
\newunicodechar{∈}{\ensuremath{\in}}
\newunicodechar{↑}{\ensuremath{\uparrow}}
\newunicodechar{¬}{\ensuremath{\neg}}
\newunicodechar{⊥}{\ensuremath{\bot}}
\newunicodechar{↝}{\ensuremath{\leadsto}}
\newunicodechar{↶}{\ensuremath{\curvearrowleft}}
\newunicodechar{↺}{\ensuremath{\circlearrowleft}}
\newunicodechar{⊔}{\ensuremath{\sqcup}}
\newunicodechar{⨆}{\ensuremath{\bigsqcup}}
\newunicodechar{∩}{\ensuremath{\cap}}
\newunicodechar{∪}{\ensuremath{\cup}}

\newunicodechar{ℓ}{\ensuremath{\ell}}

\newunicodechar{Δ}{\ensuremath{\Delta}}
\newunicodechar{Γ}{\ensuremath{\Gamma}}
\newunicodechar{Σ}{\ensuremath{\Sigma}}
\newunicodechar{Θ}{\ensuremath{\Theta}}
\newunicodechar{Ω}{\ensuremath{\Omega}}

\newunicodechar{α}{\ensuremath{\alpha}}
\newunicodechar{β}{\ensuremath{\beta}}
\newunicodechar{δ}{\ensuremath{\delta}}
\newunicodechar{ε}{\ensuremath{\varepsilon}}
\newunicodechar{φ}{\ensuremath{\phi}}
\newunicodechar{γ}{\ensuremath{\gamma}}
\newunicodechar{ι}{\ensuremath{\iota}}
\newunicodechar{κ}{\ensuremath{\kappa}}
\newunicodechar{λ}{\ensuremath{\lambda}}
\newunicodechar{μ}{\ensuremath{\mu}}
\newunicodechar{ψ}{\ensuremath{\psi}}
\newunicodechar{η}{\ensuremath{\eta}}
\newunicodechar{ρ}{\ensuremath{\rho}}
\newunicodechar{σ}{\ensuremath{\sigma}}
\newunicodechar{τ}{\ensuremath{\tau}}
\newunicodechar{ξ}{\ensuremath{\xi}}
\newunicodechar{ζ}{\ensuremath{\zeta}}
\newunicodechar{Π}{\ensuremath{\Pi}}

\newunicodechar{𝓒}{\ensuremath{\mathcal{C}}}
\newunicodechar{𝓔}{\ensuremath{\mathcal{E}}}
\newunicodechar{𝓕}{\ensuremath{\mathcal{F}}}
\newunicodechar{𝓡}{\ensuremath{\mathcal{R}}}
\newunicodechar{𝓢}{\ensuremath{\mathcal{S}}}
\newunicodechar{𝓣}{\ensuremath{\mathcal{T}}}
\newunicodechar{𝓥}{\ensuremath{\mathcal{V}}}
\newunicodechar{𝓦}{\ensuremath{\mathcal{W}}}



\begin{document}
\title{}
%\titlenote{Produces the permission block, and
%  copyright information}
%\subtitle{Extended Abstract}
%\subtitlenote{The full version of the author's guide is available as
%  \texttt{acmart.pdf} document}

\title{Generic Level Polymorphic N-ary Functions}

\author{Guillaume Allais}
\affiliation{%
  \institution{University of Strathclyde}
  \city{Glasgow}
  \country{UK}
}
\email{guillaume.allais@strath.ac.uk}

% \renewcommand{\shortauthors}{B. Trovato et al.}

\begin{abstract}
Agda's standard library struggles in various places with n-ary functions and
relations. It introduces congruence and substitution operators for functions
of arities one and two, and provides users with convenient combinators for
manipulating indexed families of arity exactly one.

After a careful analysis of the kinds of problems the unifier can easily solve,
we design a unifier-friendly representation of n-ary functions. This allows us
to write generic programs acting on n-ary functions which automatically reconstruct
the representation of their inputs' types by unification. In particular, we can
define fully level polymorphic n-ary versions of congruence, substitution and the
combinators for indexed families, all requiring minimal user input.
\end{abstract}

\maketitle

\section{Introduction}

For user convenience, Agda's standard library has accumulated a set of
equality-manipulating combinators of varying arities (Section~\ref{sec:nary2})
as well as a type-level compositional Domain Specific Language to write clean types involving
indexed families of arity exactly one (Section~\ref{sec:unarycombinators}).
None of these solutions scale well. By getting acquainted with the unifier
(Section~\ref{sec:unifier}), we can design a good representation of n-ary
function spaces (Section~\ref{sec:naryfunction}) which empowers us to write
generalised combinators (Sections \ref{sec:narycombinators} and \ref{sec:narycong})
usable with minimal user input. We then see how the notions introduced to
tackle our original motivations can be mobilised for other efforts in generic
programming from an arity-generic zipWith (Section~\ref{sec:zipwith}) to a
direct style definition of printf (Section~\ref{sec:printf}).
This paper is a literate Agda file\footnote{The source code is available at
\url{https://github.com/gallais/nary}}; we discuss some of the more
esoteric aspects of the language in Appendix~\ref{appendix:agda}.

\section{N-ary Combinators... for N up to 2}\label{sec:nary2}

Agda's standard library relies on propositional equality defined as
a level polymorphic inductive family. It has one constructor (\AIC{refl})
witnessing the fact any value is equal to itself.

\ExecuteMetaData[StateOfTheArt.tex]{equality}

As one would expect from a notion of equality, it is congruent (i.e.
for any function equal inputs yield equal outputs) and substitutive
(i.e. equals behave the same with respect to predicates). Concretely
this means we can write the two following functions by dependent
pattern-matching on the equality proof:

\ExecuteMetaData[StateOfTheArt.tex]{cong}
\ExecuteMetaData[StateOfTheArt.tex]{subst}

However we quickly realise that it is convenient to be able to use
congruence for functions that take more than one argument and
substitution for at least binary relations. The standard library
provides binary versions of both of these functions:

\ExecuteMetaData[StateOfTheArt.tex]{cong2}
\ExecuteMetaData[StateOfTheArt.tex]{subst2}

If we want to go beyond arity two we are left to either define
our own ternary, quaternary, etc. versions of \AF{cong} and \AF{subst},
or to awkwardly chain the ones with a lower arity to slowly massage
the expression at hand into the shape we want. Both of these solutions
are unsatisfactory.

\paragraph{Wish} We would like to define once and for all two functions
\AF{congₙ} and \AF{substₙ} of respective types (pseudocode):

\medskip
\noindent\begin{tabular}{@{}l@{~}l}
  \AF{congₙ} : & (\AB{f} : \AB{A₁} → ⋯ → \AB{Aₙ} → \AB{B}) →\\
               & \AB{a₁} \AD{≡} \AB{b₁} → ⋯ → \AB{aₙ} \AD{≡} \AB{bₙ} → \\
               & \AB{f} \AB{a₁} ⋯ \AB{aₙ} \AD{≡} \AB{f} \AB{b₁} ⋯ \AB{bₙ}
\end{tabular}
\medskip

\noindent\begin{tabular}{@{}l@{~}l}
  \AF{substₙ} : & (\AB{R} : \AB{A₁} → ⋯ → \AB{Aₙ} → \AF{Set} \AB{r}) →\\
                & \AB{a₁} \AD{≡} \AB{b₁} → ⋯ → \AB{aₙ} \AD{≡} \AB{bₙ} → \\
                & \AB{R} \AB{a₁} ⋯ \AB{aₙ} → \AB{R} \AB{b₁} ⋯ \AB{bₙ}
\end{tabular}

\section{Invariant Respecting Programs}

A key feature of dependently typed languages is the ability to enforce
strong invariants. Inductive families~\cite{DBLP:journals/fac/Dybjer94}
are essentially classic inductive
types where one may additionally bake in these strong invariants. As soon
as the programmer starts making these constraints explicit, they need to
write constraints-respecting programs. Although a lot of programs are
index-preserving, users need to be painfully explicit about things that
stay the same (i.e. the index being threaded all across the function's
type) rather than being able to highlight the important changes.

\subsection{Working With Indexed Families}\label{sec:unarycombinators}

The standard library defines a set of handy combinators to talk about
indexed families without having to manipulate their index explicitly.
These form a compositional type-level Domain Specific
Language~\cite{DBLP:journals/csur/Hudak96} (DSL):
each combinator has a precise semantics and putting them together builds
an overall meaning.

A typical expression built using this DSL follows a fairly simple schema:
a combinator acting as a quantifier for the index (Section~\ref{sec:quantifiers})
surrounds a combination of pointwise liftings of common type constructors
(Section~\ref{sec:liftings}), index updates (Section~\ref{sec:iupdates}),
and base predicates. This empowers us to write lighter types, which hide
away the bits that are constant, focusing instead on the key predicates
and the changes made to the index.

Before we can even talk about concrete indexed families, describe these
various combinators, and demonstrate their usefulness, we need to
introduce the data the families in our running examples will be indexed
over. We pick \AD{List} the level polymorphic type of lists parameterised
by the type of their elements: it is both well-known and complex enough
to allow us to write interesting types.

\ExecuteMetaData[StateOfTheArt.tex]{list}

The most straightforward non-trivial indexed family we can define over
\AD{List} is the predicate lifting \AF{All} which ensures that a given
predicate \AB{P} holds of all the elements of a list. It has two
constructors which each bear the same name as their counterparts in the
underlying data: nil (\AIC{[]}) states that all the elements in the
empty list satisfy \AB{P} and cons (\AIC{\_::\_}) states that \AB{P}
holds of all the elements of a non-empty list if it holds of its head
and of all the elements in its tail.

\ExecuteMetaData[StateOfTheArt.tex]{all}\label{def:all}

Some of our examples require the introduction of \AD{Any}, the other classic
predicate lifting on list. It takes a predicate and ensures that it holds of
at least one element of the list at hand. Either it holds of the first one
and we are given a proof (\AIC{here}) or it holds of a value somewhere in the
tail (\AIC{there}).

\ExecuteMetaData[StateOfTheArt.tex]{any}

\subsubsection{Quantifiers}\label{sec:quantifiers}

We have two types of quantifiers: existential and universal. As they are
meant to \emph{surround} the indexed expression they are acting upon, we
define them as essentially pairs of matching opening and closing brackets.
The opening one is systematically decorated with a mnemonic symbol: \AF{∃}
for existential quantification, \AF{Π} for explicit dependent quantification
and \AF{∀} for implicit universal quantification. Additionally we use angle
brackets for existential quantifiers and square brackets for universal ones,
recalling the operators diamond and box of modal logic.

\paragraph{Existential Quantifier}\label{def:sigma} In type theory, existential
quantifiers are represented as dependent pairs. We introduce \AR{Σ}, a dependent
record parameterised by a type \AB{A} and a type family \AB{P}. It has two fields
\ARF{proj₁} for a value of type \AB{A} and \ARF{proj₂} for a proof of type
{(\AB{P} \ARF{proj₁})}. We can build and pattern-match against pairs using the
constructor \AIC{\_,\_} and we can project either of the pair's components simply
by using its field's name.

\ExecuteMetaData[StateOfTheArt.tex]{sigma}

The existential quantifier for indexed families is defined as a special
case of \AR{Σ}; it takes the index \AF{Set} implicitly.

\ExecuteMetaData[StateOfTheArt.tex]{exists}

Using \AF{∃⟨\_⟩} we can write our first statement about an indexed family:
from the existence of a list such that \AB{P} holds of all its elements,
we can construct a list of pairs of elements and proofs that \AB{P} holds
for that value.

\ExecuteMetaData[StateOfTheArt.tex]{toList}

\paragraph{Universal Quantifiers}\label{par:universalquantifiers}
The natural counterpart of existential quantification
is universal quantification. In type theory this corresponds to a dependent
function space. Here we have room for variations and we can consider both
the explicit (\AF{Π[\_]}) \emph{and} the implicit (\AF{∀[\_]}) universal
quantifiers.

\ExecuteMetaData[StateOfTheArt.tex]{universal}
\ExecuteMetaData[StateOfTheArt.tex]{iuniversal}

Provided that a proposition holds of any value, we can prove it will hold
of any list of values by induction on such a list. Because we perform
induction on the list it is convenient to take it as an explicit argument
whereas the proof itself can take its argument implicitly.

\ExecuteMetaData[StateOfTheArt.tex]{replicate}

\subsubsection{Pointwise Liftings}\label{sec:liftings}

Pointwise liftings for an index type \AB{I} are operators turning a type
constructor on \AF{Set}s into one acting on \AB{I}-indexed families by
threading the index. They are meant to be used partially applied so that
both their inputs and their output are \AB{I}-indexed, hence the mismatch
between their arity and the number of places for their arguments.

\paragraph{Implication} We start with the most used of all: implication
i.e. functions from proofs of one predicate to proofs of another.
\ExecuteMetaData[StateOfTheArt.tex]{implies}

The combinator \AF{\_⇒\_} associates to the right just like the type
constructor for functions does. We can write the analogue of sequential
application for applicative functors~\cite{DBLP:journals/jfp/McbrideP08}
like so:
\ExecuteMetaData[StateOfTheArt.tex]{ap}

\paragraph{Conjunction} To state that the conjunction of two predicates hold
we can use the pointwise lifting of pairing.

\ExecuteMetaData[StateOfTheArt.tex]{conjunction}

This enables us to write functions which return more than one result. We can
for instance write the type of \AF{unzip}, the proof that if the conjunction
of \AB{P} and \AB{Q} holds of all the elements of a given list then both
\AB{P} and \AB{Q} in isolation hold of all of that list's elements.

\ExecuteMetaData[StateOfTheArt.tex]{unzip}

Notice that we are using the conjunction combinator both on predicates ranging
over values and on ones ranging over lists of values.

\paragraph{Disjunction} To formally describe the disjunction of two predicates,
we need to define \AD{\_⊎\_} the type of disjoint sums first. It has two
constructors each of which corresponds to a choice of one side of the sum or
the other.

\ExecuteMetaData[StateOfTheArt.tex]{sum}

The disjunction of two predicates is then the pointwise lifting of \AD{\_⊎\_}.

\ExecuteMetaData[StateOfTheArt.tex]{disjunction}

A typical use case for disjoint sums is the notion of decidability: either
a predicate or its negation holds. We can formulate a general decidability
result for \AD{All}: if for any value either \AB{P} or \AB{Q} holds then for
any list of values, either {(\AD{Any} \AB{P})} or {(\AD{All} \AB{Q})} holds.

\ExecuteMetaData[StateOfTheArt.tex]{decide}

Here we did not limit ourselves to either \AB{P} or its negation but it is
sometimes necessary to talk directly about negation.

\paragraph{Negation} Traditionally negation is defined as functions into
the empty type \AD{⊥}. We start by defining it as the inductive type with
not constructor together with its elimination principle (\AF{⊥-elim}).

\begin{minipage}[t]{0.225\textwidth}
  \ExecuteMetaData[StateOfTheArt.tex]{bot}
\end{minipage}\begin{minipage}[t]{0.225\textwidth}
  \ExecuteMetaData[StateOfTheArt.tex]{botelim}
\end{minipage}

Negation for a unary predicate \AB{P} is then the unary predicate which maps
\AB{i} to {(\AB{P} \AB{i} → ⊥)}.

\ExecuteMetaData[StateOfTheArt.tex]{negation}

The two predicate liftings \AD{All} and \AD{Any} interact in non-trivial ways.
For instance if we know that the negation of \AB{P} holds of any value in a
given list then \AB{P} can't hold of all its elements. In other words: a
single counter-example is enough to disprove a universal statement.

\ExecuteMetaData[StateOfTheArt.tex]{anynotall}

Notice that we are once more using the combinator we just defined both on
a predicate on values and one on lists of values.

\subsubsection{Adjustments To The Ambient Index}\label{sec:iupdates}

Threading the index is only the least invasive of the modes of action
available to us. But we can also more actively interact with the ambient
index either by ignoring it completely, adjusting it using a function
or overwriting it entirely. We will not detail the last option as, as always,
overwriting is adjusting with a constant function.

\paragraph{Constant} Although we have so far only manipulated indexed families,
some of our function's arguments or its result may not depend on the index.
The \AF{const}ant indexed family is precisely what we need to represent
these cases.

\ExecuteMetaData[StateOfTheArt.tex]{const}

We can for instance prove that if the constantly false predicate
{(\AF{const} \AF{⊥})} holds true of all the elements of a list then
said list is the empty list. We use a section (i.e. a partially applied
infix operator) of propositional equality to formulate that conclusion.
In the proof we do not need to consider the \AIC{\_∷\_} case: Agda
automatically detects that it is impossible.

\ExecuteMetaData[StateOfTheArt.tex]{empty}

Note that we had to add a type annotation to \AIC{[]}: the type of the
index of the predicate defined using \AF{const} is an implicit polymorphic
argument and so is the type of elements in \AD{List}'s nil constructor.
Agda can infer that these two implicit arguments are equal but needs to be
given enough information to figure out what it ought to be. In type theory,
an identity function is a fine definition of a type annotation operator:

\ExecuteMetaData[StateOfTheArt.tex]{annot}

\paragraph{Update} On the other end of the spectrum, we have operations
which update the ambient index using an arbitrary function. The notation
\AF{\_⊢\_} is inspired by the convention in type theory to consider that
proofs in sequent calculus are written in an ambient context and that we
may use a turnstile to describe the addition of newly-bound variables to
this context (see e.g. Martin L\"of's work~\cite{martin1982constructive}).

\ExecuteMetaData[StateOfTheArt.tex]{update}

Stating that a function operating on lists is compatible with \AF{All} is
a typical use case of such a combinator. If the function at hand is called
\AF{f} then the convention in the standard library is to call such a proof
\AF{f⁺} as it makes \AF{f} appear in the conclusion. We pick \AF{concat}
(whose classic definition is left out) in this concrete example.

\ExecuteMetaData[StateOfTheArt.tex]{join}

\subsection{Working With Multiple Indices}

We started by showing both the type and the implementation of each of our
examples. Although convenient at first to build an understanding of which
arguments are explicit and which ones are implicit, we are in the end only
interested in the way combinators let us write types. In this section, we
focus on the types and only the types of our examples.

The combinators presented earlier are all available in the standard library.
As we have demonstrated, they work really well for unary predicates. Unfortunately
they do not scale beyond that. Meaning that if we are manipulating binary relations
for instance we have to explicitly introduce one of the indices and partially apply
the relations in question before we can use our usual unary combinators. This leads
to cluttered types which are not much better than their fully expanded counterparts.

Let us look at an example. We introduce \AD{Pw} (for ``pointwise'') the relational
equivalent of the predicate lifting \AD{All} we have been using as our running example so
far. The inductive family \AD{Pw} is parameterised by a relation \AB{R} and ensures
its two index list are compatible with \AB{R} in a pointwise manner. If both lists
are empty then they are trivially related (\AIC{[]}); otherwise we demand that their
heads are related by \AB{R} and their tails are related pointwise (\AIC{\_∷\_}).

\ExecuteMetaData[StateOfTheArt.tex]{pointwise}

To state the relational equivalent to \AD{All}'s \AF{\_<⋆>\_} using our
combinators for unary predicates, we need to partially apply \AD{Pw} to
\AB{xs} to make it a predicate as well as explicitly use a λ-abstraction to
build the relation corresponding to the fact that \AB{R} implies \AB{S}.

\ExecuteMetaData[StateOfTheArt.tex]{appw}

Ideally we could have instead used binary version of the combinators for unary
predicates we saw earlier and have simply written:

\ExecuteMetaData[Examples.tex]{appw}

We could duplicate the definitions for unary predicates and have equivalent
combinators for binary relations however this will create two new issues.
First, the day we need a library for ternary relations we will have triplicated
the initial work. Second, we would have two sets of definitions with identical
names meaning they cannot be both imported in the same module without clashing
thus forcing users to manually disambiguate each use site.

\paragraph{Wish} We would like to define once and for all n-ary quantifiers,
pointwise lifting of common type constructors, and adjustment functions.

\section{Plan} We can start to draw out the structure of our contribution now
that we have a good idea of the current state of the art, its limitation, and
the extensions we want to see. Here are the key points we deliver:

\paragraph{Reified Types} We come up with a representation of n-ary
functions which is as general as possible: the domains should be allowed to
be different types, even types defined at different universe levels.

\paragraph{Semantics} We give a semantics taking a reified type and
computing its meaning as a \AF{Set} at some universe level. This level
also needs to be computed from the description.

\paragraph{Invertible} The representation and its semantics are unifier
friendly. That is to say that if using a combinators yields a constraint of
the form ``this type is the result of evaluating the representation of
an n-ary function type'', then Agda will be able to reconstruct the
representation and discharge the constraints without any outside help.

\paragraph{Applications} Lastly we deliver the two wishes we
formulated earlier by actually implementing the n-ary versions of \AF{cong},
\AF{subst}, and the various combinators for manipulating indexed families.

\section{Getting Acquainted With the Unifier}\label{sec:unifier}

Unification is the process by which Agda reconstructs the values of the
implicit arguments the user was allowed to leave
out~\cite{DBLP:conf/tlca/AbelP11}. It is one of the
mechanisms bridging the gap between the source program which should be
convenient for humans to read, write, and modify and the fully explicit
terms in the internal syntax.

It is important to build a good understanding of the problems the unifier
can easily solve to be able to write combinators usable with minimal user
input. Indeed if we can anticipate that an argument can be reconstructed,
we may as well make it implicit and let Agda do the work.

\paragraph{Notations} We write \AB{?a} for a
metavariable, \AB{e}[\AB{?a₁}, ⋯ ,\AB{?aₙ}] for an expression \AB{e}
containing exactly the metavariables \AB{?a₁} to \AB{?aₙ},
\AB{c} \AB{e₁} ⋯ \AB{eₙ} for the constructor \AB{c} applied to \AB{n}
expressions and {\AB{lhs} ≈ \AB{rhs}} to state a unification problem
between two expressions \AB{lhs} and \AB{rhs}.

\paragraph{Unification Tests}\label{sec:unificationtest}
We can easily trigger the resolution of
unification problems by writing unit tests in the source language. We can
force Agda to introduce metavariables by using an underscore (\AF{\_}) as a
placeholder for a subterm and use \AIC{refl} at the proof of a propositional
equality to force it to unify the two expressions stated to be equal. For
instance in the following test we force Agda to check that
{(\AB{?A} → \AB{?B})} can be unified with {(\AD{ℕ} → \AD{ℕ})}.

\ExecuteMetaData[Unifier.tex]{unifproblem}

To express problems where a single metavariable is used multiple times,
we can use a \AK{let} binder. For instance, we can indeed unify
{(\AB{?A} → \AB{?A})} with {(\AD{ℕ} → \AD{ℕ})}.

\ExecuteMetaData[Unifier.tex]{sharedunifproblem}

Whenever Agda cannot solve a metavariable by unification it is highlighted
in yellow like so: \AgdaUnsolvedMeta{\_}. Whenever Agda cannot satisfy a
unification constraint raised by the use of \AIC{refl}, it will also highlight
it in yellow like so: \AgdaUnsolvedConstraint{\AIC{refl}}.

Let us now look at the various scenarios in which it is easy for the unifier
to decide whether a constraint is satisfiable.

\subsection{Instantiation}

The simplest case the unifier can encounter is a problem of the form
{\AB{?a} ≈ \AB{e}[\AB{?a₁} ⋯ \AB{?aₙ}]} where \AB{?a} does not appear
in the list {[\AB{?a₁}, ⋯ ,\AB{?aₙ}]}. The unifier can simply instantiate
the metavariable to the candidate expression.

For instance in the following test you can see that neither the underscore
on the left-hand side nor the \AIC{refl} constructor is highlighted in yellow.
Meaning that the metavariable on the left was indeed solved (by instantiating
it to the expression on the right-hand side) and that the constraint induced
by the use of \AIC{refl} was thus satisfied. The problem itself is
under-constrained so it is not surprising that the right-hand side lights up.

\ExecuteMetaData[Unifier.tex]{instantiation}

\subsection{Constructor Headed}

The second case where the unifier can easily make progress is a unification
problem between to constructor-headed expression
{\AB{c} \AB{e₁} ⋯ \AB{eₘ} ≈ \AB{d} \AB{f₁} ⋯ \AB{fₙ}}.

\paragraph{Success} Either the constructors \AB{c} and \AB{d} match up, we learn
that \AB{m} equals \AB{n} and we can reduce the problem to unifying the constructors'
respective arguments by forming the new unification problems
{(\AB{e₁} ≈ \AB{f₁}) ⋯ (\AB{eₘ} ≈ \AB{fₙ})}.

In the following example, Agda sees that both expressions have {\_→\_} as their
head constructor, proceeds to unify \AD{ℕ} with itself on the one hand (which
succeeds because both have the same head constructor and they do not have any
arguments) and \AD{ℕ} with \AB{?A} on the other (which succeeds by instantiation).

\ExecuteMetaData[Unifier.tex]{unifconstr}

\paragraph{Failure} Or \AB{c} and \AB{d} are distinct and we can immediately
conclude that unification is impossible. We cannot write an expression in Agda
demonstrating this case as it leads to a type error in the language.
Trying to form the unification problem {\AD{ℕ} ≈ (\AB{?A} → \AB{?B})} would
raise such an error because \AD{ℕ} and {\_→\_} are distinct head constructors.

\subsection{Avoid Computations...}

In general unification problems involving computations are undecidable.
We can easily construct a total simulation function \AF{sim} for Turing
machines which takes in as arguments the code for an arbitrary program \AB{prg}
and a natural number \AB{n} and returns \AN{0} if and only if the program runs
for exactly \AB{n} steps before stopping and \AN{1} otherwise. Forming a
constraint like {\AF{sim} \AB{prg} \AB{\_} ≈ \AN{0}} is effectively asking
whether the program \AB{prg} terminates. It is clearly impossible to write a
unifier solving all problems of this form.

\subsection{... In Most Cases}\label{sec:unifiermagic}

Although unification problems involving computations will in general fail
to produce solutions, there are exceptions.

\paragraph{Disappearing Problem} The first favourable case is a Lapalissade:
stuck function applications which are guaranteed to go away in all cases of
interest to us are never a problem. This is true whenever we know that in all
use cases the concrete values at hand will allow evaluation to reveal enough
constructors for unification to succeed.

To demonstrate this phenomenon we introduce a type \AF{nary} of \AB{n}-ary
functions on natural numbers. It is parameterised by the return type of the
\AB{n}-ary function and defined by induction on \AB{n}.

\ExecuteMetaData[Unifier.tex]{nary}

In general, it is impossible to solve the unification constraint
{\AF{nary} \AB{?n} \AB{?A}} ≈ {(\AD{ℕ} → \AB{A})}. If the natural number
is not specified then \AF{nary} is stuck. And there is no hope to solve
this problem; indeed there are two solutions (\AB{?n} could be either \AN{0}
or \AN{1}) because every unary function is also a nullary symbol whose type
is a function type. As explained earlier, Agda communicates to us this failure
to solve the two metavariables passed to \AF{nary} as arguments by
highlighting them in yellow. The constructor \AIC{refl} is also highlighted as
the source of the unification constraint that could not be satisfied.

\ExecuteMetaData[Unifier.tex]{unsolved}

If the natural number argument is however a concrete value then \AF{nary}
evaluates fully and Agda is able to reconstruct \AB{A} by unification. In
the following two examples we unify {(\AD{ℕ} → \AD{ℕ})} with {(\AD{ℕ} → \AB{?A})}
on the one hand and {\AB{?A}} on the other. Both unification problems succeed
without any issue.

\begin{minipage}{0.225\textwidth}
  \ExecuteMetaData[Unifier.tex]{normalised1}
\end{minipage}\begin{minipage}{0.225\textwidth}
  \ExecuteMetaData[Unifier.tex]{normalised0}
\end{minipage}

This observation is language independent. It will directly influence our
encoding: we expect our users to only ever use our generic congruence
combinator with concrete arities. A representation defined by induction
on such a natural number would therefore work well with the unifier.

\paragraph{Invertible Problem}\label{sec:unifierinversion}
The second case in which we may encounter
unification problems involving stuck computations and still see Agda
find a solution is more language dependent but just as principled.
Whenever the stuck function is defined by a set of equations whose
right-hand sides are clearly anti-unifiable, we can invert it.

For instance if the \AF{Set} parameter to \AF{nary} is known to be
\AD{ℕ} then the right-hand side of the first equation is \AD{ℕ} and
the second's one has the shape {(\AD{ℕ} → \_)}. These two are clearly
disjoint and so Agda can invert \AF{nary} and figure out that the
arity we left out in the following example is \AN{1}.

\ExecuteMetaData[Unifier.tex]{inverted}

If we had passed {(\AD{ℕ} → \AD{ℕ})} instead as the second argument to
\AF{nary} then the two right-hand sides would not have been obviously
disjoint and Agda would have given up on trying to invert \AF{nary}.

\ExecuteMetaData[Unifier.tex]{notinverted}

These two examples tell us that we can hope to leave out a function's
arity entirely if we statically know its codomain and it has a shape
clearly anti-unifiable with the right-hand sides of our semantics of
reified function types. Note in particular that combinators acting
on relations (cf. Section~\ref{sec:unarycombinators}) are manipulating
functions whose codomain is always of the shape {(\AF{Set} \_)} which
is clearly disjoint from {(\_ → \_)}. We ought to be able to define
their \AB{n}-ary counterparts without having to mention \AB{n} explicitly.

\section{Representing N-ary Function Types}\label{sec:naryfunction}

Now that we understand how the unifier works, we can design a generic
representation and its semantics (called \AF{⟦\_⟧} here for convenience)
so that whenever we have a constraint of the form
{\AF{⟦} \AB{?r} \AF{⟧} ≈ (\AD{ℕ} → \AF{Set})}, it can easily
lead to the reconstruction of the representation \AB{?r}.

\paragraph{User Input} As we have just seen, a binary function type
{(\AB{A} → \AB{B} → \AB{C})} with codomain \AB{C} can also be seen as
a unary function type with codomain {(\AB{B} → \AB{C})}. As a consequence
in the general case there is no hope to get Agda to reconstruct the
representation we have in mind without passing it at least a little bit
of information. The least we can do is tell Agda the arity of the function.
From this single natural number we will compute the shape of the whole
representation.

\paragraph{Unification} As we have just seen, if we want Agda to reconstruct
the representation from a unification constraint then our best hope is that
the semantics function evaluates fully and simply disappears. This means in
particular that it should not get stuck on a pattern-matching analysing the
representation. This can be achieved with certainty by constraining our
representation to only be built up from things we either will not pattern-match
against (e.g. \AF{Set}s) or type constructors which enjoy η-equality (i.e.
for which values can always be made to look like they are in canonical form).
Just like the representation, the semantics will have to be computed entirely
from the natural number corresponding to the function's arity.

From these two observations, we decide that our representation will be
parameterised by a natural number which we will use to compute a number
of right-nested products.

\paragraph{Right-Nested Products} The two basic building blocks of right-nested
products are a binary product \AF{\_×\_} and the unit type \AF{⊤}.

We obtain the binary product as the non-dependent special case of \AR{Σ} we
introduced in Section~\ref{def:sigma}. We did not mention it at the time but
record types in Agda enjoy η-rules. That is to say that any value \AB{p} of type
{(\AR{Σ} \AB{A} \AB{P})} is definitionally equal to
{(\ARF{proj₁} \AB{p} \AIC{,} \ARF{proj₂} \AB{p})}.

The unit type is defined as a record with no field. Every value of type
\AF{⊤} is equal to the canonical value \AIC{tt}.

\ExecuteMetaData[StateOfTheArt.tex]{unit}

Even though \AF{⊤} is defined as a \AF{Set}, we will sometimes need to use
it at a higher level. The usual solution is to manually lift it to the
appropriate level. Because \AR{Lift} is also a record, it will not get in the
way of reconstruction.

\ExecuteMetaData[StateOfTheArt.tex]{lift}

\paragraph{Level Polymorphism} To achieve fully general level polymorphism, we
need all the domains of our function type to be potentially at different levels.
Luckily the notion of \AF{Level} in Agda is a primitive \AF{Set} and we can thus
manipulate them just like any other values. In particular we can define containers
storing them. Our first definition called \AF{Levels} defines an \AB{n}-tuple of
\AF{Level}s by induction on \AB{n}.

\ExecuteMetaData[N-ary.tex]{levels}

\paragraph{Heterogeneous Domains} Before we can generate the big right-nested
\AB{n}-tuple packaging the function's domains, we need to compute the level at
which it is going to live. The definition of \AR{Σ} makes clear that the product
of two types living respectively at level \AB{a} and \AB{b} sits at level
{(\AB{a} ⊔ \AB{b})} i.e. the least upper bound of \AB{a} and \AB{b}.
We define \AF{⨆} as the generalisation of the least upper bound operator
to {(\AF{Levels} \AB{n})} by induction on \AB{n}.

\ExecuteMetaData[N-ary.tex]{tolevel}

Knowing that {(\AF{Set} \AB{a})} sits at level {(\AF{suc} \AB{a})}, it is
natural to declare that our \AB{n}-tuple of sets defined at various \AF{Levels}
will be defined at the successor of the generalised least upper bound of these
\AF{Levels}.

\ExecuteMetaData[N-ary.tex]{sets}

We can now encode an \AB{n}-ary function space as essentially a collection \AB{ls}
of {(\AF{Levels} \AB{n})} together with a corresponding \AB{n}-tuple of type
{(\AF{Sets} \AB{n} \AB{ls})} for the domains, and a level \AB{r} and a
{(\AF{Set} \AB{r})} for the codomain.

\paragraph{Semantics}\label{sec:semantics}
This encoding has a straightforward semantics by induction on \AB{n} and case
analysis on the {(\AF{Sets} \AB{n} \AB{ls})} argument. A \AIC{zero}-ary function
type is simply the codomain whilst a {(\AIC{suc} \AB{n})}-ary one is a unary
function type whose codomain is the \AB{n}-ary function type obtained by
induction hypothesis.

\ExecuteMetaData[N-ary.tex]{arrows}

If we look carefully at this definition we can notice that the function
\AF{Arrows} may only ever get stuck if the natural number is not concrete.
Even though we do take the \AF{Sets} argument apart, it is a product type
and thus enjoys η-rules. We have achieved the degree of unifier-friendliness
we were aiming for.

Our first example is a \AN{2}-ary function: our favourite indexed family \AD{All}.
The last element of the telescope, a value whose type is a lifted version of the
unit type, can be inferred by Agda so we leave it out.

\ExecuteMetaData[Examples.tex]{all}

\section{Combinators for Indexed Families}\label{sec:narycombinators}

Now that we have our generic representation of \AB{n}-ary function types, we can
finally start building the \AB{n}-nary counterparts of the combinators we
discussed at length in Section~\ref{sec:unarycombinators}.

\subsection{Quantifiers}

If we already know how to quantify over one variable, we can easily describe how
to quantify over \AB{n} variables by induction over \AB{n}. This is what \AF{quantₙ}
does. Provided a (level polymorphic) quantifier \AB{Q} and a \AF{Set}-valued
\AB{n}-ary function \AF{f}, we distinguish two cases: if \AB{n} is \AN{0} then
the function is already a \AF{Set} and we can return it directly; otherwise we use
\AB{Q} to quantify over the outer variable which we call \AB{x} and proceed to
quantify over the remaining variables in {(\AB{f} \AB{x})} by using the induction
hypothesis.

\ExecuteMetaData[N-ary.tex]{quantify}

We can define the specific instances of \AB{n}-ary quantification we are interested
in by partially applying \AF{quantₙ} with the appropriate concrete quantifiers.
Because we are dealing with \AF{Set}-valued functions, we can leave their arity as
an implicit argument and let Agda infer it at use site. In all cases we give them
the same name as their unary counterparts as they can be used as drop-in replacements
for them.

We start with the \AB{n}-ary existential quantifier defined using the unary
quantifier we introduced in Section~\ref{sec:quantifiers}.

\ExecuteMetaData[N-ary.tex]{ex}

Similarly we can define the explicit and implicit universal quantifiers.

\ExecuteMetaData[N-ary.tex]{all}
\ExecuteMetaData[N-ary.tex]{iall}

\subsection{Pointwise Liftings}

Pointwise lifting of a binary function can be defined uniformly for any
operation of type {(\AB{A} → \AB{B} → \AB{C})} and any pair of \AB{n}-ary
functions whose domains match and codomains are respectively \AB{A} and
\AB{B}. It is defined by induction on the arity \AB{n} of the input functions.

\ExecuteMetaData[N-ary.tex]{lift2}

From this very general definition we can recover the combinators we are
used to. For each one of them we are able to leave out the arity argument
thanks to the observation we made in Section~\ref{sec:unifiermagic}: \AF{Set}
and {(\AB{?A} → \AB{?B}) are anti-unifiable and Agda is therefore able to
reconstruct the arity for us!

Implication is the lifting of the function space.

\ExecuteMetaData[N-ary.tex]{implication}

Conjunction is the lifting of pairing.

\ExecuteMetaData[N-ary.tex]{conjunction}

Disjunction is the lifting of the sum type.

\ExecuteMetaData[N-ary.tex]{disjunction}

Negation is obviously not a binary operation. In practice, rather than
having multiple ad-hoc lifting functions for various arities we have a
fully generic \AF{liftₙ} functional which lifts a \AB{k}-ary operator
to work with \AB{k} \AB{n}-ary functions whose respective codomains
match the domains of the operator. Its type could be summarised as:

\medskip
\noindent\begin{tabular}{@{}l@{~}l}
  \AF{liftₙ} : & ∀ \AB{k} \AB{n}.\\
               & (\AB{B₁} → ⋯ → \AB{Bₖ} → \AB{B}) →\\
               & (\AB{A₁} → ⋯ → \AB{Aₙ} → \AB{B₁}) →\\
               \multicolumn{2}{c}{\vdots} \\
               & (\AB{A₁} → ⋯ → \AB{Aₙ} → \AB{Bₖ}) →\\
               & (\AB{A₁} → ⋯ → \AB{Aₙ} → \AB{B})\\
\end{tabular}
\medskip

The thus generalised definition has a fairly unreadable type so we leave
this formal definition out of the paper. Curious readers can consult the
accompanying code. We can evidently use \AF{liftₙ} with \AB{k} equal to
\AN{1} to lift negation from an operation on \AF{Set} to an operation on
\AF{Arrows}.

\ExecuteMetaData[N-ary.tex]{negation}

\subsection{Adjustments To The Ambient Indices}

We now have obtained the generalised versions of the index-threading combinators
we wanted. We can similarly define a number of index-altering combinators. The
first two are the \AB{n}-ary versions of the two operators we described in
Section~\ref{sec:iupdates}.

Lifting a mere value to a constant \AB{n}-ary function is a matter of composing
\AF{const} with itself \AB{n} times.

\ExecuteMetaData[N-ary.tex]{const}

Updates are a bit more subtle: now that we are not limited to a single index,
we can choose which index should be updated. We expect the user to provide a
natural number \AB{n} to target a specific index, the type of the combinator
then clearly states that \AB{n} sets are skipped, the target is updated and
the rest of the type is unchanged.

\ExecuteMetaData[N-ary.tex]{compose}

The added complexity of working with \AB{n}-ary relations means that we have
more interesting operators than simply the generalised version of the ones
we had introduced for unary predicates.

We may for instance want to map a unary function on the result of an \AB{n}-ary
one. Note that this empowers us to partially apply any \AB{n}-ary function to a
value \AB{x} in its \AB{k}-th argument by choosing to see it as a \AB{k}-ary
function and mapping {(\AF{\_\$} \AB{x})} on it.

\ExecuteMetaData[N-ary.tex]{map}

\section{Congruence and Substitution}\label{sec:narycong}

So far the types we have ascribed our combinators for \AB{n}-ary relations
were fairly tame. Things get a bit more complicated when dealing with
congruence and substitution: we will not be able to write these functions' types
directly. Both definitions follow the same structure: we start by computing
the operation's type by induction and we can then implement the operation itself.

\subsection{Congruence}

The type of congruence mentions only one function. However it is applied to two
distinct lists of values to form the left-hand side and the right-hand side of
the conclusion. As a consequence when we compute the type we take two functions
as inputs and use one to apply to the arguments meant for the left-hand side
and the other for the ones meant for the right-hand side of the equation.

Congruence for two \AN{0}-ary functions collapses to simply propositional equality
of the two constant values.

Congruence for two (\AIC{suc} \AB{n})-ary functions \AB{f} and \AB{g} amounts to
stating that for any pair of equal values \AB{x} and \AB{y} we expect that
{(\AB{f} \AB{x})} and {(\AB{g} \AB{y})} are congruent.

\ExecuteMetaData[Applications.tex]{Cong}

The congruence lemma is then obtained by stating that the \AB{n}-ary function \AB{f}
is congruent with itself. We prove it by induction on \AB{n}, pattern-matching on the
proofs of equality as we go along.

\ExecuteMetaData[Applications.tex]{cong}

\subsection{Substitution}

The definition of \AF{Substₙ} is identical to that of \AF{Congₙ} except that we
now consider predicates rather than arbitrary functions. The base case is therefore
dealing with \AB{P} and \AB{Q} being two \AF{Set}s rather than two values at a given
type. As a consequence we demand a function transporting proofs of \AB{P} to proofs
of \AB{Q} rather than a proof of equality.

\ExecuteMetaData[Applications.tex]{Subst}

Substitution acts on \AB{n}-ary relations. Recalling our observation made
in Section~\ref{sec:unifiermagic} that Agda can easily reconstruct the arity
of \AF{Set}-valued functions, we can make \AB{n} an implicit argument.

\ExecuteMetaData[Applications.tex]{subst}

\section{Further Generic Programming Efforts}

The small language we have developed to talk about \AB{n}-ary functions can
be used beyond our first few motivating examples of congruence, substitution,
and combinators to define types involving relations. We detail in this section
various results that fall out naturally from this work. We start with generic
currying and uncurrying, and then use these to define an \AB{n}-ary \AF{zipWith}
and revisit \AF{printf} in direct style.

\subsection{Product and (Un)Currying}

We gave in Section~\ref{sec:semantics} a semantics to our reified types as
proper \AB{n}-ary function types. We can alternatively interpret a \AF{Sets}
as a big right-nested and \AR{⊤}-terminated product containing one value for
each \AF{Set}. We once more proceed by induction on \AB{n}.

\ExecuteMetaData[N-ary.tex]{product}

We can convert back and forth between a unary function whose domain is a
\AF{Product} of \AF{Sets} and an \AB{n}-ary function whose domains are the
same sets. These conversion functions correspond to currying and uncurrying.
Both \AF{curryₙ} and \AF{uncurryₙ} are implemented by structural induction
on \AB{n} and in terms of their binary counterparts. In the base case, the
function is either applied to \AIC{tt} or uses \AF{const} to throw away a
value of type \AR{⊤}; this is an artefact of the fact our definition of
\AF{Product} is \AR{⊤}-terminated

\ExecuteMetaData[N-ary.tex]{curry}
\ExecuteMetaData[N-ary.tex]{uncurry}

\paragraph{\AR{⊤}-free Variant} In practice users do not tend to write
\AR{⊤}-terminated right-nested products. As a consequence it is convenient
to have a definition of \AF{Product} which has a special case for \AF{Sets}
of size exactly \AN{1} returning the \AF{Set} without pairing it with \AR{⊤}.
This makes \AF{curryₙ} and \AF{uncurryₙ} more useful overall. Most generic
functions however are easier to implement using the \AR{⊤}-terminated version
of \AF{Product}. In our library we provide both as well as conversion
functions between the two interpretations.

\subsection{N-ary Zipping Functions}\label{sec:zipwith}

Some functions are easier to write curried but nicer to use uncurried.
This is the case with \AF{zipWithₙ}, the \AB{n}-ary version of the
higher-order function which takes a function and two lists as inputs
and produces a list by processing both lists in lockstep and using the
function it was passed to combine their elements. Using ellipses, we would
write its type as:

\medskip
\noindent\begin{tabular}{@{}l@{~}l}
  \AF{zipWithₙ} : & ∀ \AB{n}. (\AB{A₁} → ⋯ → \AB{Aₙ} → \AB{B}) →\\
                  & \AD{List} \AB{A₁} → ⋯ → \AD{List} \AB{Aₙ} → \AD{List} \AB{B}
\end{tabular}
\medskip

To formally write this type, we need to explain how to map a level polymorphic
endofunctor on \AF{Set} (here: \AD{List}) over a {(\AF{Sets} \AB{n} \AB{ls})}.
We proceed by induction on \AB{n}.

\ExecuteMetaData[N-ary.tex]{smap}

As explained earlier it is vastly easier to implement the function using
the uncurried type, and to then recover the desired type by invoking generic
(un)currying in the appropriate places. The function we want is therefore
implemented in term of an auxiliary definition called \AF{zw-aux}.

\ExecuteMetaData[N-ary.tex]{zipWith}

\paragraph{Implementation} The auxiliary definition is still a bit involved
so we detail each equation of its definition below. We start with its type
first.

\ExecuteMetaData[Applications.tex]{zw-aux-type}

When \AB{n} is \AN{0}, a Haskeller would typically return an infinite list
containing the value \AB{f} repeated. However this is not possible in Agda,
a total language~\cite{DBLP:journals/jucs/ATurner04}: all the lists have to
be finite. Our only principled option is to return the empty list.

\ExecuteMetaData[Applications.tex]{zw-aux0}

Because the behaviour of the \AN{0} case is less than ideal, we bypass it
every time except if \AF{zipWithₙ} is explicitly called on \AN{0}. This is
done by having a special case for \AB{n} equals \AN{1}. In this situation,
we can get our hands on a function \AB{f} of type
{((\AB{A} \AR{×} \AR{⊤}) → \AB{R})} and a {\AD{List} \AB{A}} and we need to
return a {\AD{List} \AB{R}}. We map a tweaked version of the function on the list.

\ExecuteMetaData[Applications.tex]{zw-aux1}

The meat of the definition is in the last case: we are given a function
{((\AB{A} \AR{×} \AB{A₀} \AR{×} ⋯ \AR{×} \AB{Aₙ}) → \AB{R})}, a list of \AB{A}s
and a product of lists {(\AD{List} \AB{A₀} \AR{×} ⋯ \AR{×} \AD{List} \AB{Aₙ})}.
We massage the function to obtain another one of type
{((\AB{A₀} \AR{×} ⋯ \AR{×} \AB{Aₙ}) → (\AB{A} → \AB{R}))} which we can
combine with the product of lists thanks to our induction hypothesis. This gives
us back a list of functions of type {(\AB{A} → \AB{R})}. We can conclude thanks
to the usual binary \AF{zipWith} to combine this list of functions with the list
of arguments we already had.

\ExecuteMetaData[Applications.tex]{zw-auxn}

\subsection{Printf}\label{sec:printf}

The combinators we have introduced also make it easy to implement \AF{printf}
in direct style as opposed to the classic accumulator-based
definition~\cite{DBLP:conf/icfp/Augustsson98,DBLP:journals/jfp/Danvy98}. We
effectively produce a well typed version of the ill typed intermediate
function Asai, Kiselyov, and Shan consider in their derivation of a direct-style
implementation using delimited control~\cite{DBLP:journals/lisp/AsaiKS11}.

We work in a simplified setting which allows us to focus on the contribution our
\AB{n}-ary combinators bring to the table. Our \AF{printf} will only take natural
numbers as arguments and we will not worry about defining the lexer transforming
a raw \AD{String} into a \AF{Format}, that is to say a list of \AD{Chunk}s each
being either a \AIC{Nat} corresponding to a ``\%u'' directive (i.e. unsigned
decimal integer) or a \AIC{Raw} string.

\begin{minipage}[t]{0.25\textwidth}
  \ExecuteMetaData[Printf.tex]{chunk}
\end{minipage}\begin{minipage}[t]{0.2\textwidth}
  \ExecuteMetaData[Printf.tex]{chunks}
\end{minipage}

Our notion of \AF{Format} is not intrinsically sized but we do need to know
how many arguments our \AF{printf} function is going to take if we want to use
the machinery for \AB{n}-ary functions. We assume the existence of a \AF{size}
function counting the number of \AIC{Nat} in a \AF{Format}. We also assume the
existence of \AF{0ℓs}, a (\AF{Levels} \AB{n}) equal to \AF{0ℓ} everywhere. Using
these we can give \AF{Format} a semantics as a \AF{Sets} of arguments \AF{printf}
will expect. To each \AIC{Nat} we associate a \AD{ℕ} constraint, the other
\AD{Chunk}s do not give rise to the need for an input.

\ExecuteMetaData[Printf.tex]{format}

The essence of \AF{printf} is then given by a function \AF{assemble} which
collects a list of strings from various sources. Whenever the format expects
a natural number, we know we got one as an input and can \AF{show} it.
Otherwise the raw string to use is specified in the \AF{Format} itself as
an argument to \AIC{Raw}.

\ExecuteMetaData[Printf.tex]{assemble}

The toplevel function is obtained by currying the composition of
\AF{concat} and \AF{assemble}.

\ExecuteMetaData[Printf.tex]{printf}

We can check on an example that we do get a function with the appropriate
type when we use a concrete \AF{Format} (here the one we would obtain from
the string \AStr{"\%u < \%u"}).

\ExecuteMetaData[Printf.tex]{example}

And that it does produce the expected string when run on arguments.

\ExecuteMetaData[Printf.tex]{test}

\section{Conclusion, Related and Future Work}

We have seen that Agda's standard library defines a useful couple of functions
to produce proofs of equality as well as a type-level domain specific language
to manipulate unary predicates. We then got acquainted with the unifier and the
process by which a unification constraint can lead to the reconstruction of a
function's implicit arguments. Based on this knowledge we have designed a
representation of \AB{n}-ary function types particularly amenable to such
reconstructions. This allowed us to define \AB{n}-ary versions of congruence,
substitution as well as the basic building blocks of the type-level DSL for
relations we longed for. The notions introduced to set the stage for these
definitions were already powerful enough to allow us to revisit classic
dependently typed traversals such as an \AB{n}-ary version of \AF{zipWith},
or direct-style \AF{printf}.
This work has now been merged in the Agda standard library and will be
part of the released version 1.1 (see modules \texttt{Data.Product.Nary.NonDependent},
\texttt{Function.Nary.NonDependent}, and \texttt{Relation.Nary} for
library code and \texttt{Text.Printf} for one application).

\paragraph{Limitations} We are relying heavily on two key features of Agda
that are not implemented in other dependently typed languages as far as we know.

First, \AF{Level}s are a first class notion: they can be stored in
data structures, passed around in functions and computed with just like
any other primitive type. Unlike other primitive types (e.g. floating point
numbers), it is not possible to perform case analysis on them. Other
dependently typed languages may not be too keen on adopting this extension
given that its meta-theoretical consequences are currently unknown. We
recommend that they use meta-programming instead to duplicate this work:
programs written in MetaCoq~\cite{DBLP:conf/itp/AnandBCST18,draf/metacoq18}
can for instance explicitly manipulate universe levels.

Second, Agda's unifier has a heuristics that attempts to invert stuck functions
when solving constraints. As we have explained in Section~\ref{sec:unifierinversion}
this heuristics is principled: if it succeeds, the generated solution is
guaranteed to be unique. We hope that our detailed use-case incites other
languages to consider adopting it.

\paragraph{Codes for N-ary Function Types} We can find in the literature various
deep~\cite{DBLP:conf/icfp/VerbruggenVH08} and
shallow~\cite{DBLP:journals/jfp/McBride02,DBLP:conf/plpv/WeirichC10}
embeddings of polymorphic types and a fortiori of \AB{n}-ary function types in
a dependently typed language. However none of them are fully level polymorphic
and most only consider the representation as a secondary requirement, their
focus being on certifying equivalent programs in Generic Haskell~\cite{DBLP:conf/popl/Hinze00}.
We however care deeply about level polymorphism as well as being unification-friendly
to minimise the reification work the user needs to do.

\paragraph{Telescopes} The lack of dependencies between the various domains
and the codomain of our \AF{Arrows} is flagrant. A natural question to ask
is how much of this machinery can be generalised to telescopes rather than
mere \AF{Sets} without incurring any additional burden on the user. From
experience we know that it is sometimes wise to explicitly use the
non-dependent version of an operator (e.g. function composition) to inform
Agda's unifier that it is only looking for a solution in a restricted
subset.

\paragraph{Datatype genericity} Our implementation of an \AB{n}-ary version of
\AF{zipWith} started as an example of the types and accompanying generic programs one can
write with our library. It demonstrates that the notions introduced for our
purposes can be useful in a more general context. This result is not new either
with or without dependent
types~\cite{DBLP:journals/jfp/FridlenderI00,DBLP:journals/jfp/McBride02}.
It can however be extended as previous efforts in dependently typed programming
have demonstrated: Weirich and Casinghino's work~\cite{DBLP:conf/plpv/WeirichC10}
on arity-generic but also data-generic programming suggests we should be
able to push this further. Their development predates the addition of universe
polymorphism to Agda and although the traversals are adequately heterogeneous,
their approach would not scale to universe polymorphic functions.

\paragraph{Parametricity as a derivation principle} Some of the examples we have
used could have been obtained ``for free'' by parametricity: \AD{All} and \AD{Pw}
are respectively the predicate and the relational inductive-style translations of
the definition of \AD{List}. The function \AF{replicate} defined in
Section~\ref{par:universalquantifiers}
is a consequence of the abstraction theorem corresponding to the predicate
interpretation and a similar free theorem stating that if a relation is reflexive
then so is its pointwise lifting could have been derived for \AD{Pw}. Bernardy,
Jansson, and Paterson's work on parametricity for dependent
types~\cite{DBLP:journals/jfp/BernardyJP12} makes these observations formal and
generalises these constructions to all inductive types and n-ary relational
liftings.

\appendix
\section{Agda-Specific Features}\label{appendix:agda}

We provide here a description of some of the more esoteric Agda
features used in this paper. Readers interested in a more thorough
introduction to the language may consider reading Ulf Norell's lecture
notes~\cite{DBLP:conf/afp/Norell08}.

\subsection{Syntax Highlighting in the Text}

The colours used in this paper all have a meaning: keywords are highlighted
in \AK{orange}; \AF{blue} denotes function and type definitions; \AIC{green}
marks constructors; \ARF{pink} is associated to record fields and
corresponding projections.

\subsection{Universe Levels}\label{appendix:agda:level}
\balance

Agda avoids Russell-style paradoxes by introducing a tower of universes
\AF{Set₀} (usually written \AF{Set}), \AF{Set₁}, \AF{Set₂}, etc. Each
\AF{Setₙ} does not itself have type \AF{Setₙ} but rather \AF{Setₙ₊₁} thus
preventing circularity.

We can form {\bf function types} from a domain type in \AF{Setₘ} to a codomain
type in \AF{Setₙ}. Such a function type lives at the level corresponding
to the maximum of \AB{m} and \AB{n}. This maximum is denoted {(\AB{m} \AF{⊔} \AB{n})}.

An {\bf inductive} type or a {\bf record} type storing values of type \AF{Setₙ}
needs to be defined at universe level \AB{n} or higher. We can combine multiple
constraints of this form by using the maximum operator. The respective definitions
of propositional equality in Section~\ref{sec:nary2} and dependent pairs in
Section~\ref{def:sigma} are examples of such data and record types.

Without support for a mechanism to define {\bf level polymorphic functions},
we would need to duplicate a lot of code. Luckily Agda has a primitive notion
of universe levels called \AD{Level}. We can write level polymorphic code by
quantifying over such a level \AB{l} and form the universe at level
\AB{l} by writing (\AF{Set} \AB{l}). The prototypical example of such a level
polymorphic function is the identity function \AF{id} defined as follows.

\ExecuteMetaData[Appendix.tex]{identity}

\subsection{Meaning of Underscore}

Underscores have different meanings in different contexts. They can either stand
for argument positions when defining identifiers, trivial values Agda should be
able to reconstruct, or discarded values.

\subsubsection{Argument Position in a Mixfix Identifier}

Users can define arbitrary mixfix identifiers as names for both functions and
constructors. Mixfix identifiers are a generalisation of infix identifiers
which turns any alternating list of name parts and argument positions into a
valid identifier~\cite{DBLP:conf/ifl/DanielssonN08}. These argument positions
are denoted using an underscore. For instance \AF{∀[\_]} is a unary operator,
(\AIC{\_::\_}) corresponds to a binary infix identifier and (\AF{\_\%=\_⊢\_}) is a
ternary operator.

\subsubsection{Trivial Value}

Programmers can leave out trivial parts of a definition by using an underscore
instead of spelling out the tedious details. This will be accepted by Agda as
long as it is able to reconstruct the missing value by unification. We discuss
these use cases further in Section~\ref{sec:unificationtest}.

\subsubsection{Ignored Binder}

An underscore used in place of an identifier in a binder means that the binding
should be discarded. For instance {(λ \_ → a)} defines a constant function.
Toplevel bindings can similarly be discarded which is a convenient way of
writing unit tests (in type theory programs can be run at typechecking time)
without polluting the namespace. The following unnamed definition checks for
instance the result of applying addition defined on natural numbers to
\AN{2} and \AN{3}.

\ExecuteMetaData[Appendix.tex]{unittest}

\subsection{Implicit Variable Generalisation}\label{appendix:agda:variable}

Agda supports the implicit generalisation of variables appearing in type
signatures. Every time a seemingly unbound variable is used, the reader can
safely assume that it was declared by us as being a \AK{variable} Agda should
automatically introduce. These variables are bound using an implicit
prenex universal quantifier. Haskell, OCaml, and Idris behave similarly with
respect to unbound type variables.

In the type of the following definition for instance, \AB{A} and \AB{B} are
two \AF{Set}s of respective universe levels \AB{a} and \AB{b} (see
Appendix~\ref{appendix:agda:level}) and \AB{x} and \AB{y} are two values of
type \AB{A}. All of these variables have been introduced using this implicit
generalisation mechanism.

\ExecuteMetaData[StateOfTheArt.tex]{cong}

If we had not relied on the implicit generalisation mechanism, we would have
needed to write the following verbose type declaration.

\ExecuteMetaData[Appendix.tex]{congtype}

This mechanism can also be used when defining an inductive family.
In Section~\ref{def:all}, we introduced the predicate lifting \AD{All}
in the following manner. The careful reader will have noticed a
number of unbound names: \AB{a}, \AB{A}, \AB{p} in the declaration
of the type constructor and \AB{x} and \AB{xs} in the declaration of
the data constructor \AIC{\_::\_}.

\ExecuteMetaData[StateOfTheArt.tex]{all}

This definition corresponds internally to the following expanded
version (modulo the order in which the variables have been generalised
over).

\ExecuteMetaData[Appendix.tex]{all}

\section*{Acknowledgements}

We would like to thank the reviewers for their helpful comments and their
suggestions to discuss parametricity as a derivation principle, and to add
an appendix to make the paper accessible to a wider audience.

The research leading to these results has received funding from EPSRC
grant EP/M016951/1.

\bibliographystyle{ACM-Reference-Format}
\bibliography{nary}

\end{document}